\documentclass[amsmath,amssymb,prb,
showpacs,twocolumn,
superscriptaddress,
prl]{revtex4-1}
\usepackage{graphicx,bm,color,subfigure}
\usepackage[T1]{fontenc}
\usepackage{mathrsfs}
\usepackage{bm}
\usepackage{amsmath}
\usepackage{amssymb}
\usepackage{graphicx}
\usepackage{esint}
\usepackage{multirow}
\usepackage{float}
\usepackage{array}
\usepackage{makecell}
\usepackage{harpoon}
\usepackage{booktabs}
\usepackage{gensymb}
\usepackage{simplewick}
\usepackage{subfigure}
\usepackage{soul}
\usepackage{ulem}


\begin{document}
\title{The magnetic $Z_2$ topological insulator on the AA-stacked bilayer graphene}
\author{Yu-Bo Liu}
\thanks{These two authors contributed equally to this work.}
\affiliation{School of Physics, Beijing Institute of Technology, Beijing 100081, China}
\author{Zhi-Yan Shao}
\thanks{These two authors contributed equally to this work.}
\affiliation{School of Physics, Beijing Institute of Technology, Beijing 100081, China}
\author{Ye Cao}
\email{ycao@bit.edu.cn}
\affiliation{School of Physics, Beijing Institute of Technology, Beijing 100081, China}
\author{Fan Yang}
\email{yangfan_blg@bit.edu.cn}
\affiliation{School of Physics, Beijing Institute of Technology, Beijing 100081, China}
\date{\today}
\begin{abstract}
The properties displayed by graphene at van Hove singularities (VHS) have caught significant attention in recent years. The emergence of exotic quantum states at these singularities prompts investigations on their evolution within the realm of multilayer stacking structures. In our research, we delve into the study of a repulsive Hubbard model focusing on the AA-stacked bilayer graphene at VHS.
Within the system's ground state, each of the top and bottom layers hosts a set of spin-density waves (SDWs). These SDWs each takes on three mutually perpendicular spin polarization directions. Importantly, there is noteworthy feature that their spin polarization directions in the two layers exist as elegant embodiments of antiferromagnetic arrangement, persvading the structure with a striking pattern. Referred to in prior research as the chiral SDWs, this intralayer density wave structure confers the system the characteristics of a Chern topological insulator.
However, what is particularly fascinating is the pure divergence of the bilayer structure's topological traits when compared to its monolayer counterpart. The system exhibits a profound symmetry known as $Z_2$, preserving its invariance under the combined operations of time-reversal and interlayer exchange. Consequentely, the system's ground state manifests a seemingly trivial Chern number, yet harbors a profound and intricate nontrivial  $Z_2$ topological invariant. These remarkable observations align our findings with the conceptual framework of the quantum spin Hall effect.
\end{abstract}

\maketitle

\section{\bf Introduction}\label{Introduction}
In the field of condensed matter physics, the discovery of graphene~\cite{Novoselov2004}, a single layer of carbon atoms arranged in a two-dimensional (2D) honeycomb lattice, has enriched the theoretical world. 
Particularly, the Dirac cones at specific points in the Brillouin zone~\cite{Neto2009} induces a unique 2D Dirac operator, which behaves as if it has an effective zero mass and a speed of light $c/300$~\cite{Novoselov2005}. This physical phenomena is inspiring. These electronic correlations can become significant at the Van Hove singularity (VHS), a point of high electronic density of states (DOS), which is opposite to the customary consideration that they are weak. At this singularity, under quarter electron doping, the Fermi surface nesting phenomenon takes place, leading to the emergence of three spin density waves (SDWs) with spin polarization directions that are mutually perpendicular~\cite{Li2012}, similar to its triangular lattice counterpart~\cite{Martin2008, Kato2010}.
Because of non-zero Chern number, these chiral states behave as Chern topological insulators, unique phases of matter that are electrically insulating in their bulk but support conducting states on their surfaces due to their nontrivial topological order and illustrate quantum anomalous Hall (QAH) effect~\cite{Haldane1988, Sticlet2013, Wang2017, Chang2022}. 

The exploration for multilayer graphene materials, typically because of the properties of monolayer graphene, have shown ever increasing theoretical and experimental interest, to name a few~\cite{Latil2006, Nilsson2006, LopesdosSantos2007, Koshino2007, Andres2008, Nakamura2008, Min2008, Nilsson2008, Yan2009, Liu2009, Min2009, Koshino2010, McCann2010, Koshino2011, Mele2011, Henriksen2012, Sakurai2012, Cappelluti2012, Rakhmanov2012, Lin2012, Paton2014, Kindermann2015,GarciaMartinez2015, Campos2016, Rozhkov2016, Wu2017, Bao2017, Xu2018, Nam2018, Mora2019, Liu2019, Horii2019, Zhao2020, Mohammadi2021, Li2022, Wang2022, Ledwith2022,  Lu2022, Wang2024}. The AA-stacked form, where all layers are perfectly aligned, the Bernal AB-stacked form, and the intriguing twisted layers are the components of the main configurations for bilayer graphene. In the AB-stacked configuration, half of the atoms lie over an atom of the previous layer, while the other half lie directly over the center of a hexagon in the previous layer. The twisted bilayer graphene is a completely different form, achieved by twisting one layer relative to the other at a specific angle. This configuration is unique and the relvevent electronic properties has turned out to be exceptional and tunable , thus we have seen exciting possibilities for various applications~\cite{caoyuan20181,caoyuan20182}. 

There is scientific consensus that the quantum states depend on the arrangement of layers, and the bonds between consecutive layers are much weaker than those between neighboring atoms within the same layer.
Generally, the AB-stacked structure is considered more stable than the AA-stacked one, resulting in a larger body of theoretical and experimental research focused on the former~\cite{Koshino2007, Nakamura2008, Min2008, Nilsson2008, Yan2009, Min2009, Koshino2010, McCann2010, Koshino2011, Sakurai2012, Cappelluti2012, GarciaMartinez2015, Bao2017, Horii2019}. However, the latter exhibits greater symmetries and is commonly found in various scenarios, winning increasing interest~\cite{ Andres2008,Min2008, Min2009, Liu2009, Lin2012,Rozhkov2016, Bao2017, Mohammadi2021}.

While getting deeper insight into the topological insulators, certain questions receive serious attention: does the AA-stacked form of bilayer graphene possess nontrivial topological states, and if so, whether or not these states accurately match the current classification of topological states?
After reviewing relationship between the Kane-Mele model~\cite{Kane2005-1, Kane2005-2} and the Haldane model~\cite{Haldane1988}, wherein the SO term in the Kane-Mele model induces two sets of Haldane models, then these two sets fall together due to the Rashba effect, thereby resulting in a system with two sets of energy bands having opposite Chern numbers, defined as a new $Z_2$ topological state with a Chern number of zero and exhibiting quantum spin Hall (QSH) effect~\cite{GarciaMartinez2015}, we set out to investigate whether similar correlations exist within our bilayer system.

Before specific calculations, we undertake a thought experiment: a natural correlation carring both the QAH effect and QSH effect stands out because of the stacking monolayer graphene in the AA arrangement with SDW order. In this scenario, the chiral SDW order in each layer induces a QAH state, analogous to a Haldane model. The interlayer coupling parallels with the Rashba term in the Kane-Mele model, effectively connecting two equivalent Haldane models. Should the final ground state display opposite chiralities between the top and bottom layers, this model would reproduce the QSH effect.

In this study, we extend the principles of the Kane-Mele model to AA-stacked bilayer graphene, described by a repulsive Hubbard model with weak interlayer hopping interactions. To maintain consistency, we logically consider that the SDW configuration in each layer mirrors that of the monolayer structure. Our main focus centers around understanding the relative arrangement of the two sets of SDW orders between the layers. 
Adopting the Ginzburg-Landau (G-L) theory, we carefully restrict all possible configurations to two discrete scenarios. We then perform numerical computations to optimize the energy for each configuration. After a thorough comparison of the results, we identify the ground state of the system, wherein the chiralities of the top and bottom layers are opposite while their polarization directions exhibit antiparallel behavior for the aligned vectors in two layers. 
As a result, the Chern number of the ground state on bilayer graphene is zero, which is a simple addition of the Chern numbers of the two layers, 1 + (-1) = 0. Despite this, the system presents doubly degenerate boundary states that cross the energy gap. The existence of these boundary states is protected by the combined time-reversal and layer exchange symmetries. This joint symmetry, together with an additional 2D inversion symmetry, is responsible for the degeneracy of these boundary states. In this situation, we further employ two methods to calculate the $Z_2$ topological invariant. One approach is based on constructing the product of $\mathrm{U}(2N)$ Berry connection along the Wilson loop~\cite{Yu2011}, and the other involves a specialized treatment using the parity of occupied Bloch wave functions in systems with inversion symmetry~\cite{Fu2007}. Both methods yielded nontrivial $Z_2$ topological invariants, and the results were consistent.

In conclusion, our team come up with the achievement of a $Z_2$ topological state in a bilayer structure without spin-orbit coupling. Using the methods proposed by Fu and Kane~\cite{Fu2007} and Yu et al.~\cite{Yu2011} to calculate the $Z_2$ topological invariant, we identify the ground state as a magnetic $Z_2$ topological insulator. In this system, the two opposite chiralities can be analogously interpreted as two distinct spin states. The interplay between these states, arising from the presence of interlayer tunneling, can be naturally likened to the Kane-Mele model~\cite{Kane2005-1,Kane2005-2}, where the coupling of two sets of QAH states with different spins generates the QSH state. This discovery deepens our understanding of the complex topological properties of bilayer graphene.

The rest of the paper is organized as follows. In Sec.~\ref{Hamiltonian}, we first review the chiral SDW state of the monolayer system. We reproduce this ground state using the G-L method in Appendix A, demonstrating the effectiveness and power of the G-L method. Then the Hamiltonian and mean-field Hamiltonian of the bilayer system is presented.
In Sec.~\ref{GL}, utilizing the broken symmetries of the system, we employ the G-L method to obtain two possible ground-state density wave configurations.
In Sec.~\ref{topo}, through numerical computations, we ultimately determine the ground state of the system and establish its nontrivial $Z_2$ topological nature. Additionally, we analyze the existence and degeneracy of boundary states based on symmetry considerations.
Finally, Sec.~\ref{conclusion} is devoted to the conclusions and outlooks.

\section{\bf The Model and Hamiltonian of bilayer graphene}\label{Hamiltonian}
\begin{figure}[htbp]
\centering
\includegraphics[width=0.48\textwidth]{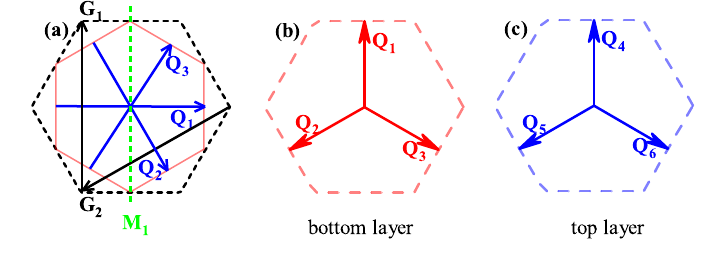}
\caption{(Color online) (a) The hexagonal Fermi surface (depicted in red) is inscribed within the Brillouin zone (shown as a black dotted line) of graphene at the Van Hove singularity (VHS). The vectors $\mathbf{Q}_p$ (with $p=1,2,3$) represent nesting vectors, whereas $\mathbf{G}_1$ and $\mathbf{G}_2$ denote the elementary reciprocal vectors. The green dotted line symbolizes a mirror axis of the system, denoted by $\mathrm{M}_1$. When there is weak interlayer hopping, the configuration of the wave vectors in the bottom (b) and top (c) layers is preserved and depicted in a manner that could vary by an integral multiple of reciprocal lattice vectors from the configuration in (a).}\label{vec_q}
\end{figure}

Before delving into bilayer graphene, we provide a concise overview of the ground state characteristics of its monolayer counterpart. As illustrated in Fig.~\ref{vec_q}(a), the graphene lattice (considering only nearest-neighbor hopping) exhibits a delightful phenomenon known as perfect Fermi surface nesting at the VHS, occurring at a quarter electron doping. This gives rise to three nested vectors: $\mathbf{Q}_1=\mathbf{G}_1/2$, $\mathbf{Q}_2=\mathbf{G}_2/2$, and $\mathbf{Q}_3=-(\mathbf{G}_1+\mathbf{G}_2)/2$, where $\mathbf{G}_1$ and $\mathbf{G}_2$ denote elementary reciprocal vectors. By employing the repulsive Hubbard model on monolayer graphene, we arrive at the mean-field Hamiltonian for the intriguing SDW as follows:
\begin{eqnarray}\label{Eq:mean field monolayer}
H_{\text{mf}} &=& \sum_{\langle\mathbf{ij}\rangle\sigma}-\tilde{t}_{ij}(c_{\mathbf{i}\sigma}^{\dagger}c_{\mathbf{j}\sigma}+h.c.)\nonumber\\
&&+ \beta\sum_{\mathbf{k} p\alpha s_{1,2} }\chi_{p}^{\alpha}c_{\mathbf{k} s_1}^{\alpha\dagger}c_{\mathbf{k}+\mathbf{Q}_p s_2}^{\alpha}(\bm{\sigma}\cdot\mathbf{n}_p)_{s_1 s_2},
\end{eqnarray}
where $-\tilde{t}_{\mathbf{ij}} =-t-\mu\delta_{\mathbf{ij}}$ is the hopping term $t$ combined with chemical potential $\mu$, $\alpha\in(A, B)$ is the sublattice index and $\chi_{p}^{\alpha}=\pm 1$, $\beta=2Um/3\sqrt{3}$, with $m$ denoting the amplitude of the order parameter and $U$ representing the strength of the repulsive Hubbard interaction. Here, $\mathbf{Q}_p$ (with $p=1,2,3$) corresponds to the aforementioned SDW wave vector shown in Fig.~\ref{vec_q}(b), while $\mathbf{n}_{p}$ ($p=1,2,3$) are arbitrary 3D unit vectors that represent the spin polarization direction of the three SDWs.

In previous studies~\cite{Li2012}, the primary objective was to determine the polarization direction of the three SDWs. Researchers relied on rigorous numerical calculations presented in their paper~\cite{Li2012} to obtain the lowest energy state and confirm that these SDWs exhibit mutually perpendicular polarization directions, i.e. $\mathbf{n}^{1,2,3}$ in Eq.~\eqref{Eq:mean field monolayer} correspond to the $\mathbf{e}_{x,y,z}$ components, respectively, while $\chi_{p}^{A}=1(p=1,2,3)$ and $\chi_{1}^B=\chi_{2}^B=-\chi_{3}^B=-1$. This intriguing chiral SDW state can be comprehensively described as a Chern topological insulator, featuring a nontrivial Chern number. As a point of reference, we compare these findings to the well-known G-L theory, which provides further insights into the directions of these SDW polarizations without relying on numerical computations (see appendix in \ref{sdw}). Applying the G-L theory helps us strengthen the understanding of the configuration of SDWs and validate our results.

Based on previous studies of SDW in monolayer graphene, this paper focuses on investigating AA-stacked bilayer graphene at a quarter electron doping level. The analysis considers only the nearest neighbor hopping for both interlayer and intralayer hoppings. The repulsive Hubbard model for the bilayer graphene is expressed as follows:

\begin{eqnarray}\label{hubbard_bilayer}
H &=& \sum_{\langle\mathbf{ij}\rangle\sigma\nu}-\tilde{t}_{\mathbf{ij}}(c_{\mathbf{i}\sigma}^{\nu\dagger}c_{\mathbf{j}\sigma}^{\nu}+h.c)+ U\sum_{\mathbf{i}\nu}c_{\mathbf{i}\uparrow}^{\nu\dagger}c_{\mathbf{i}\uparrow}^{\nu}c_{\mathbf{i}\downarrow}^{\nu\dagger}c_{\mathbf{i}\downarrow}^{\nu} \nonumber\\
&& +t'\sum_{\mathbf{i}\sigma}(c_{\mathbf{i}\sigma}^{t\dagger}c_{\mathbf{i}\sigma}^{b}+h.c).
\end{eqnarray}
Here,  $\nu =\text{t},\text{b}$ labels the top and bottom layer respectively. The third term represents the nearest interlayer hopping, with $t'= 0.05t$ considered as much smaller than $t$.
Since the interlayer hopping is much smaller than the intralayer hopping, we regard the interlayer hopping as a perturbation, and the interlayer hopping will not seriously destroy the Fermi surface nesting of each layer. There are still three SDWs with perpendicular spin polarization direction in the bottom and top layers respectively, but the relative spin polarization direction of the two layers needs to be further determined. 

In the mean-field treatment, the Hubbard model on bilayer graphene lattice has the form,
\begin{eqnarray}\label{Eq:SDW_possible_state}\nonumber
	H_{mf} &=&\sum_{\langle\mathbf{ij}\rangle\sigma\nu}-\tilde{t}_{\mathbf{ij}}(c_{\mathbf{i}\sigma}^{\nu\dagger}c_{\mathbf{j}\sigma}^{\nu}+h.c.)+t'\sum_{\mathbf{i}\sigma}(c_{\mathbf{i}\sigma}^{t\dagger}c_{\mathbf{i}\sigma}^{b}+h.c) \\  \nonumber
	&&+ \beta\sum_{\mathbf{k} p\alpha s_1 s_2}\chi_{p}^{\alpha}c_{\mathbf{k} s_1}^{b\alpha\dagger}c_{\mathbf{k}+\mathbf{Q}_p s_2}^{b\alpha}(\bm{\sigma}\cdot\mathbf{n}_p)_{s_1 s_2} \\
	&&+ \beta\sum_{\mathbf{k} q\alpha s_1 s_2}\chi_{q}^{\alpha}c_{\mathbf{k} s_1}^{t\alpha\dagger}c_{\mathbf{k}+\mathbf{Q}_q s_2}^{t\alpha}(\bm{\sigma}\cdot\mathbf{n}_q)_{s_1 s_2}, \\\nonumber
	\end{eqnarray}
where  $p\in\{1,2,3\}$ and $q\in\{4,5,6\}$ are two indices that respectively index the three wave vectors of SDWs for the top and bottom layer. $\alpha\in(A, B)$ is the sublattice index and $\chi_{p}^{A}=1 (p\in{1,2,...,6})$, $\chi_{1}^{B}=\chi_{2}^{B}=-\chi_{3}^{B}=-1$ and $\chi_{4}^{B}=\chi_{5}^{B}=-\chi_{6}^{B}=-1$. $\mathbf{n}_p (p=1,2,3)$ and $\mathbf{n}_q (q=4,5,6)$ represent the spin polarization directions of the three SDWs in the bottom and top layers, respectively.
As mentioned earlier, in the case where interlayer tunneling is weak, we can conclude that the wave vector configurations of the SDWs for the top and bottom layers (as shown in Fig.~\ref{vec_q}(b) and (c)) as well as the relative spin polarization directions of the intralayer SDWs are the same as those in their respective monolayer structures. What we need to focus on is how the relative spin polarization directions between the two sets of density waves will arrange when considering the interactions in the bilayer structure.

\section{\bf G-L analysis}\label{GL}

In the density wave states of the bilayer system, there are six independent spin polarization directions of density wave oscillations. To obtain the ground state, one could treat these directions as arbitrary and optimize the variational parameters to find the configuration with the lowest energy. However, a more efficient approach involves using G-L theory combined with the system's symmetry to classify these directions into discrete possible configurations. By optimizing the energy for each specific configuration and comparing the final results, we can determine the ground state. This method significantly reduces the number of variational parameters and alleviates the numerical computation burden.

The original Hamiltonian possesses multiple symmetries. The Hubbard model itself exhibits time-reversal, translation, and spin SU(2) symmetries. On the honeycomb lattice, it is additionally endowed with $\mathrm{D}_6$ point group symmetry. Furthermore, the AA-stacked bilayer structure possesses interlayer exchange symmetry. The system's free energy, which is a function of the spin polarization directions \{$\mathbf{n}$\} where \{$\mathbf{n}$\} is short for $\{\mathbf{n}_{1},\cdots,\mathbf{n}_6\}$ denoting six directions in the Eq.~\eqref{Eq:SDW_possible_state}, must remain unchanged under these symmetry operations.
These symmetry operations on the mean-field Hamiltonian in Eq.~\eqref{Eq:SDW_possible_state} is equivalent to \{$\mathbf{n}$\} $\rightarrow$ \{$\mathbf{n}'$\}. 

It is obervious  that the time-reversal operation dictates $\{\mathbf{n'}\} =-\{\mathbf{n}\}$. 
The point group $\mathrm{D}_6$ includes two generators: The $\pi/3$  counterclockwise rotation $\mathrm{C}_6$ and mirror symmetry $\mathrm{M}_1$, the operations of them result in
${\mathbf{n'}}_{1(4)}={\mathbf{n}}_{2(5)}$, ${\mathbf{n'}}_{2(5)}={\mathbf{n}}_{3(6)}$, ${\mathbf{n'}}_{3(6)}={\mathbf{n}}_{1(4)}$ and
${\mathbf{n}'}_{1(4)}={\mathbf{n}}_{1(4)}$, ${\mathbf{n}'}_{2(5)}={\mathbf{n}}_{3(6)}$, ${\mathbf{n}'}_{3(6)}={\mathbf{n}}_{2(5)}$ respectively.
The mean-field Hamiltonian of the system breaks the translational symmetry in three density wave vector directions. Consequently, we require the free energy to remain invariant under these three translational operations. Each of these operations causes the polarization direction of one density wave to reverse while leaving the other two unchanged. Since the free energy remains invariant under point group rotations, the results induced by these three operations can be obtained by combining any one of them with the $\mathrm{C}_6$ rotational operation. Hence, we only need to consider one of them for our analysis. Let the two primitive lattice vectors be denoted as $a_1$ and $a_2$. In real space, the SDW term $c_{\mathbf{k} s_1}^{bA\dagger}c_{\mathbf{k}+\mathbf{Q}_p s_2}^{bA}$ takes the form of $e^{i\mathbf{Q}_p\cdot \mathbf{R}}c_{\mathbf{R} s_1}^{bA\dagger}c_{\mathbf{R} s_2}^{bA}$. Translating by $a_1$ will lead to $\mathbf{n'}_{1(4)}=-\mathbf{n}_{1(4)}$, $\mathbf{n'}_{2(5)}=\mathbf{n}_{2(5)}$, $\mathbf{n'}_{3(6)}=-\mathbf{n}_{3(6)}$.
In addition, the spin SU(2) symmetry requires the free energy to contain only dot product terms such as $\mathbf{n}_p\cdot \mathbf{n}_{p'}$. The interlayer exchange symmetry does not impose any additional constraints in this context.

\begin{figure}[htbp]
	\centering
	\includegraphics[width=0.5\textwidth]{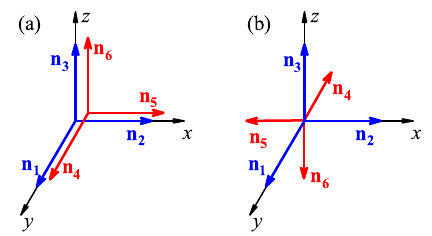}
	\caption{(Color online) The candidate configurations of spin polarization directions \{$\mathbf{n}$\} of SDWs for the ground state. (a) describe the ground state when $\gamma < 0$,  whereas (b) belongs to the ground state when $\gamma > 0$. }\label{Fig:configurations}
\end{figure}

Since the interlayer hopping is much smaller, there are still three SDWs with perpendicular spin polarization direction in the bottom and top layers, respectively. Considering all the mentioned symmetries, the expression of the free energy takes the following form: 
\begin{eqnarray}\label{freefunction3}
\mathcal{F} =\gamma m^2(\mathbf{n}_1\cdot\mathbf{n}_4+\mathbf{n}_2\cdot\mathbf{n}_5+\mathbf{n}_3\cdot\mathbf{n}_6)+o(m^4),
\end{eqnarray}
where $\gamma$ is an arbitrary real numbers and $m$ is the amplitude of order parameters. 
As a result, the sign of $\gamma$ leads to two distinct configurations for the ground state of the system. On the one hand, when $\gamma$ is negative, the spin polarization directions of the density waves with the same wavevectors in the top and bottom layers are parallel to each other, as depicted in Fig.~\ref{Fig:configurations}(a). On the other hand, if $\gamma$ is positive, the spin polarization directions of these density waves in the top and bottom layers are antiparallel to each other, as illustrated in Fig.~\ref{Fig:configurations}(b).

In summary, through G-L analysis, we have identified two possible density wave configurations (Fig.~\ref{Fig:configurations}(a) and (b)). The specific configuration that the system will adopt will be determined through mean-field optimization.

\section{\textbf{Topological Property}}\label{topo}
The study of topological properties of the ground state plays a pivotal role in understanding the nature of a given physical system, and the topological classification~\cite{Fidkowski2010,Morimoto2015,Chiu2016,Kruthoff2017} of these states has emerged as a central focus in the field of condensed matter physics. This classification, which is deeply intertwined with the global properties of the system, can reveal unique phases of matter and predict their novel quantum behaviors. Topological states usually refer to a state which is gapped in the bulk but metallic at the boundary, however the classification using topological invariants is different. 

Upon examining the two configurations shown in the Fig.~\ref{Fig:configurations}, two distinct states based on the chirality of the density waves in the top and bottom layers can be discerned. In configurations (a), the chirality is identical in both layers, while in configurations (b), the chirality is opposite.
A Chern topological state is inherent within a monolayer structure, leading these two states to different topological classifications. For the former state, the total topological invariant (Chern number) is nontrivial as it arises from identical invariants in each layer. In contrast, the latter state's total Chern number is inevitably zero due to the invariants of the two layers being of opposite values.
Therefore, if the final ground state configuration is of the first type, the topological classification of the bilayer structure aligns with that of a monolayer, except the doubled Chern number. However, if the final ground state aligns more with the second state, it results in a topological classification distinct from the monolayer structure.
Our approach involves conducting energy optimization for each of the two unique density wave configurations in Fig.~\ref{Fig:configurations}. This process enables us to pinpoint the configuration that yields the most energetically favorable outcome.

Constrained by the decomposition form of the SDW, the mean-field energy can be expressed according to Wick's theorem as follows:
\begin{eqnarray}\label{mf_energy}
E &=& \left\langle-t\sum_{\langle\mathbf{ij}\rangle\sigma\nu}c_{\mathbf{i}\sigma}^{\nu\dagger}c_{\mathbf{j}\sigma}^{\nu}+t'\sum_{\mathbf{i}\sigma}c_{\mathbf{i}\sigma}^{t\dagger}c_{\mathbf{i}\sigma}^{b}\right\rangle+c.c\nonumber\\
&+& U\sum_{\mathbf{i}\nu} \left\langle c_{\mathbf{i}\uparrow}^{\nu\dagger}c_{\mathbf{i}\uparrow}^\nu\right\rangle \left\langle c_{\mathbf{i}\downarrow}^{\nu\dagger}c_{\mathbf{i}\downarrow}^\nu\right\rangle-\left\langle c_{\mathbf{i}\uparrow}^{\nu\dagger}c_{\mathbf{i}\downarrow}^\nu\right\rangle\left\langle  c_{\mathbf{i}\downarrow}^{\nu\dagger}c_{\mathbf{i}\uparrow}^\nu\right\rangle.
\end{eqnarray}
In the above discussion, the symbol $\langle\cdots\rangle$ represents a selected SDW mean-field state. Remarkably, for $U/t$ values of $3\sim6$, we consistently observe that the energy corresponding to the configuration shown in Fig.~\ref{Fig:configurations}(b) is lower, indicating that this configuration becomes the ground state of the system for large interactions. This outcome is intuitively sensible since the repulsive Hubbard model tends to demonstrate antiferromagnetic order in the presence of large $U$. Consequently, a plausible scenario emerges where the formation of a bilayer structure leads to two layers of density waves with opposite polarization direction (chirality).
Finally, the SDW Hamiltonian is illustrated as
\begin{eqnarray}\label{Eq:SDW_ground_state}\nonumber
	H_{mf} &=&\sum_{\langle\mathbf{ij}\rangle\sigma\nu}-\tilde{t}_{\mathbf{ij}}(c_{\mathbf{i}\sigma}^{\nu\dagger}c_{\mathbf{j}\sigma}^{\nu}+h.c.)+t'\sum_{\mathbf{i}\sigma}(c_{\mathbf{i}\sigma}^{t\dagger}c_{\mathbf{i}\sigma}^{b}+h.c) \\  \nonumber
	&&+ \beta\sum_{\mathbf{k} p\alpha s_1 s_2}\chi_{p}^{\alpha}c_{\mathbf{k} s_1}^{b\alpha\dagger}c_{\mathbf{k}+\mathbf{Q}_p s_2}^{b\alpha}(\bm{\sigma}\cdot\mathbf{n}_{p})_{s_1 s_2} \\
	&&- \beta\sum_{\mathbf{k} q\alpha s_1 s_2}\chi_{q}^{\alpha}c_{\mathbf{k} s_1}^{t\alpha\dagger}c_{\mathbf{k}+\mathbf{Q}_q s_2}^{t\alpha}(\bm{\sigma}\cdot\mathbf{n}_{q})_{s_1 s_2}, \\\nonumber
	\end{eqnarray}
where $\mathbf{n}_{1(4)}=\mathbf{e}_x, \mathbf{n}_{2(5)}=\mathbf{e}_y, \mathbf{n}_{3(6)}=\mathbf{e}_z$. The other parameters are the same with those in Eq.~\eqref{Eq:SDW_possible_state}.
\begin{figure}[htbp]
\centering
\includegraphics[width=0.48\textwidth]{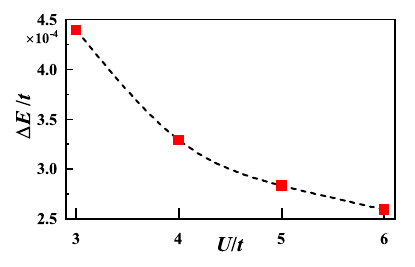}
\caption{(Color online) The energy difference $\Delta E=E_a-E_b$ between configurations (a) and (b) shown in Fig.~\ref{Fig:configurations} calculated at $U/t=3$ to $6$, where $E_{a}$ and $E_b$ represent the energy per site of configurations (a) and (b), respectively.}\label{energy}
\end{figure}

\begin{figure}[htbp]
	\centering
	\includegraphics[width=0.48\textwidth]{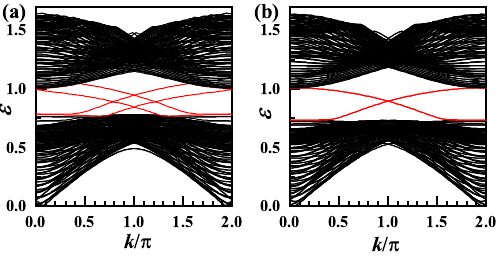}
	\caption{(Color online) (a) and (b) illustrate the boundary states (red line) of the state described in Fig.~\ref{Fig:configurations}(a) and the ground state, respectively. Notably, each red line in (b) represents two degenerate boundary states, as deduced from symmetry analysis. Throughout the calculation, we have set $U=5t$, $\epsilon$ is in units of $t$, and the width of the open boundary direction is 40 lattice constants for both (a) and (b).}\label{Fig:z1}
\end{figure}

By employing mean-field calculations, we have established that the ground state of this system possesses a Chern number of zero. Nevertheless, what makes this finding significant is that despite having a Chern number of zero, the ground state is far from topologically trivial.
In this context, we first examine the boundary states for the configuration in Fig.~\ref{Fig:configurations}(b) that corresponds to the ground state of our system, as shown in Fig.~\ref{Fig:z1}(b). For comparative analysis, the boundary states of the state represented in Fig.~\ref{Fig:configurations}(a) are illustrated in Fig.~\ref{Fig:z1}(a). Despite the Chern number of the ground states being zero, the boundary states persist and span the entire gap. 

The key to understanding the outcome lies in the structure of the ground state, which exhibits a profound combined symmetry arising from both time-reversal and interlayer exchange. This symmetry leads to a unique characteristic where a boundary state $|\mathbf{k},t\rangle$ can yield another boundary state $|-\mathbf{k},b\rangle$ upon operation with time-reversal + layer exchange, both states having degenerate eigenvalues, where $t(b)$ represents the top(bottom) layer. Consequently, the quadratic degeneracy of the boundary states at $\mathbf{k}=\pi$ is safeguarded by the system's time-reversal + layer exchange symmetry, substantiating the existence of the boundary state. Consequently, we can infer that the ground state is a topological insulator, distinguishing it from conventional insulators.

It is vital to emphasize that the ground state features doubly degenerate boundary states, a property preserved by the system's inherent symmetry. In the monolayer system, there are two boundary states, and any weak tunneling between the layers has only minimal impact, resulting in four boundary states in the bilayer system. Let's designate the top and bottom layers as $t$ and $b$, and use $r$ and $l$ to denote the left and right boundaries in the 2D configuration.
In the absence of interlayer coupling, the top and bottom layers each host two boundary states, denoted as $\psi_{r, t}$, $\psi_{l,t}$, $\psi_{l, b}$, and $\psi_{r,b}$. When introducing small interlayer tunneling, these four states experience slight changes, but their labels remain applicable.
Under a 2D inversion operation, $\psi_{l,t}(\mathbf{k})$ transforms into $\psi_{r,t}(-\mathbf{k})$, while under the joint operation of time-reversal and layer exchange, $\psi_{l,t}(\mathbf{k})$ becomes $\psi_{l,b}(-\mathbf{k})$. The symmetry of the density wave Hamiltonian under 2D inversion and time-reversal + layer exchange ensures that $\psi_{r,t}(-\mathbf{k})$ and $\psi_{l,b}(-\mathbf{k})$ possess identical energy levels, thus confirming their degeneracy. Likewise, the same analysis applies to $\psi_{l,t}$ and $\psi_{r,b}$.

In the presence of a combined symmetry resulting from time-reversal and interlayer exchange, the topological classification of the system's density wave states becomes more evident and comprehensible.
Drawing from the Kane-Mele model~\cite{Kane2005-1,Kane2005-2} that proposed the $Z_2$ topological insulator protected by time-reversal symmetry, we find a parallel in our ground state. Our system, while similar to the Kane-Mele model, is protected by time-reversal + layer exchange symmetry. To further investigate this parallel, we calculated the $Z_2$ topological invariant using two different methods.

The first method proposed by Rui Yu et al.~\cite{Yu2011} involves calculating the evolution of Wannier centers in the Brillouin zone to gain direct insights into the system's topological nature. Wannier centers are geometric properties of wavefunctions in reciprocal space, and their evolution reveals the system's topological characteristics. To compute the Wannier center, we need to construct the Wilson loop. The eigenstates of the ground state are denoted as $|\mathbf{k},\alpha\rangle$, where $\mathbf{k}=(k_x,k_y)$ represents a 2D momentum, and $\alpha$ is the band index. In our calculations, there are 32 bands, with 20 bands occupied at the VHS.

\begin{figure}[htbp]
\centering
\includegraphics[width=0.48\textwidth]{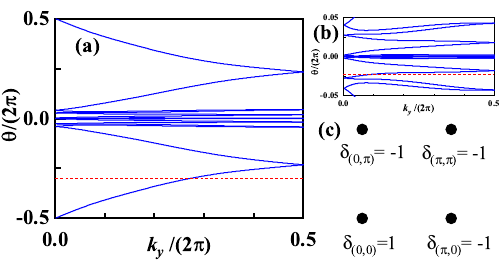}
\caption{(Color online) (a) The evolution lines of Wannier centers (blue line) with an auxiliary (red dotted) line running parallel to the $X$-axis. In (b), we provide a magnification of (a) for a closer view. (c) illustrates the product of the parity eigenvalues of the occupied states $\delta_i$ at four time-reversal momentum points.}\label{Fig:z2}
\end{figure}

For each $k_y$, we express the overlapping matrix $F_{i,i+1}(k_y)$ as follows:
\begin{equation}\label{fky}
F_{i,i+1}^{\alpha\beta}(k_y)=\langle k_y,k_{x,i},\alpha|k_y,k_{x,i+1},\beta\rangle,
\end{equation}
where $k_{x,i}=2\pi i/(aL)$ (for $i=1,2,...,L$), $a$ is the lattice constant, and $L$ is the number of unit cells in the x direction. The symbols $\alpha$ and $\beta$ represent the occupied bands. By multiplying all of the $F_{i,i+1}(k_y)$ matrices in sequence, we obtain the matrix $D(k_y)$:
\begin{equation}\label{dky}
D(k_y)=\prod_{i=1}^L F_{i,i+1}(k_y).
\end{equation}
The Wannier centers $\theta(k_y)$ are given by the phases of all the eigenvalues of the matrix $D(k_y)$. Analyzing the evolution lines of Wannier centers as $k_y$ ranges from 0 to $\pi$, we can calculate the $Z_2$ topological invariants. When the evolution of the Wannier center encloses an odd (even) number of times in the Brillouin zone as $k_y$ ranges from $0$ to $\pi$, the system is topologically nontrivial (trivial). This method provides a robust means to calculate the $Z_2$ topological invariant, avoiding the gauge-fixing problems associated with earlier methods.
The analysis is as follows: any auxiliary line parallel to the $X$-axis intersects these evolution lines an odd or even number of times. For an odd number of intersections, the $Z_2$ topological invariant is unity, indicating that the system is topologically nontrivial. Conversely, for an even number of intersections, the $Z_2$ topological invariant is zero, indicating that the system is topologically trivial. As shown in Fig.~\ref{Fig:z2}(a) and (b), the auxiliary line always intersects with the evolution lines of Wannier centers an odd number of times, indicating that the system is a $Z_2$ topological insulator.

The second method proves to be highly effective for systems that possess additional $Z_2$ symmetry, besides time-reversal symmetry. It was originally introduced by Liang Fu and C. L. Kane in Ref.~\cite{Fu2007} for systems exhibiting inversion symmetry.
This method leverages the presence of inversion symmetry to simplify the computation of $Z_2$ topological invariants. In the presence of inversion symmetry, the $Z_2$ invariants can be determined based on the parity of occupied Bloch wave functions at time-reversal invariant points in the Brillouin zone. This approach facilitates the identification of specific materials that may exhibit topological insulating behavior, providing a more accessible means to calculate $Z_2$ topological invariants in systems with inversion symmetry.

In our system, the aforementioned 2D inversion symmetry is still maintained, and the calculation approach almost remains the same. Specifically, $(0,0),(0,\pi),(\pi,0)$, and $(\pi,\pi)$ are four time-reversal + interlayer exchange invariant momenta. At these points, we calculate the product of the parity eigenvalues of all the occupied states, denoted as $\delta_i$, as depicted in Fig.\ref{Fig:z2}(c). The $Z_2$ topological invariant $v$ of the system is expressed as~\cite{Fu2007}:
\begin{eqnarray}\label{z2}
(-1)^v=\prod_i \delta_i.
\end{eqnarray}
Fig.\ref{Fig:z2}(c) confirms that the $Z_2$ topological invariant is 1, indicating that the ground state is $Z_2$ topologically nontrivial, and the results from both methods are consistent.

To summarize, it is the combined symmetry of time-reversal and layer exchange that garant the ground state's topological properties, with the spin polarization directions between two layers exhibiting antiferromagnetic ordering. Consequently, the ground state represents a magnetic $Z_2$ topological insulator.
\section{\bf Conclusion}\label{conclusion}
In summary, we have unraveled the scenario of the QSH effect in a bilayer AA-stacked graphene. Our approach involved a combination of symmetry analysis and G-L treatment, which enabled discretization of the ground state configuration for the system. Through energy optimization, we conclusively determined that the ground state consists of two sets of SDWs with opposite chirality between the top and bottom layers.
By performing calculations of the Chern number and the $Z_2$ invariant, we successfully classified this state as a $Z_2$ topological state. Additionally, we delved into the underlying reasons for the existence of degenerate boundary states through a comprehensive analysis of the system's symmetry.
We propose that establishing a connection between the two opposite chiralities and interlayer tunneling in this system, and the spin states and Rashba term in the Kane-Mele model, elegantly maps the concept of this system to the Quantum Spin Hall state.

In addition, further investigations on the ground state of the system, which exhibits fascinating physical properties, could be fruitful. By considering terms that break the time-reversal + layer exchange symmetry, such as disorder, we may observe the emergence of a small energy gap in the boundary state, leading to the appearance of corner states\cite{Ahn2019}. These corner states could potentially be indicative of a higher-order topological insulator.

\section*{Acknowledgements}
We are thankful to Sheng-Shan Qin and Cheng-Cheng Liu for discussions.
This research was supported by the NSFC under the Grant Nos. 12074031, 12234016, 12174024, 11704029, and 11674025.

\appendix
\section{G-L analysis for monolayer graphene}\label{sdw}
For the monolayer system, the Fermi surface nesting at the Van Hove singularity induces the ground state to have three SDWs with wave vectors $\mathbf{Q}_{1,2,3}$, as shown in Fig.~\ref{vec_q}, while their spin polarization directions are uncertain. In the mean-field treatment, the Hubbard model becomes:
\begin{eqnarray}\label{ham_mf_mono}
H_{mf} &=& H_{\mathrm{tb}} + \beta\sum_{\mathbf{k} p \alpha s_1 s_2}\chi_{p}^\alpha c_{\mathbf{k} s_1}^{\alpha\dagger}c_{\mathbf{k}+\mathbf{Q}_p s_2}^{\alpha}(\bm{\sigma}\cdot\mathbf{n}_p)_{s_1 s_2},\nonumber\\
\end{eqnarray}
where $\alpha\in(A, B)$ is the sublattice index, and $\chi_{p}^{\alpha}=\pm 1$, $H_{\mathrm{tb}}$ is the tight-binding Hamiltonian, and other terms are consistent with those in Eq.~\eqref{Eq:mean field monolayer}. $\mathbf{n}_{1,2,3}$ are three unit vectors representing the spin polarization directions for the corresponding ordered vector.

The density wave state breaks the system's time-reversal, $\mathrm{D}_6$ point group, spin SU(2), and translation symmetries along the three density wave vector directions. Therefore, these corresponding operations are premised on symmetries of the system's free energy. The action of these operations on the mean-field Hamiltonian can be equivalent to transformations on the three SDW polarization directions, and their specific forms are detailed in the preceding paragraph of Eq.~\eqref{freefunction3} in the main body.
In this system, the second-order free energy is trivial and does not provide any constraints. Considering the fourth-order free energy, the only admissible expression that satisfies all symmetries is:
\begin{equation}\label{freefunction1}
\mathcal{F}^{(4)} = \gamma'm^4 [(\mathbf{n}_1\cdot\mathbf{n}_2)^2 + (\mathbf{n}_1\cdot\mathbf{n}_3)^2 + (\mathbf{n}_2\cdot\mathbf{n}_3)^2].
\end{equation}
Clearly, if $\gamma'$ is positive, then these three polarization directions are mutually orthogonal. Conversely, if $\gamma'$ is negative, then they are collinear.
By performing numerical computations and optimizing the energy of these two discrete configurations, it can be determined that the former state is the ground state of the system.

\end{document}